\documentclass[11pt,a4paper]{article}
\pdfoutput=1
\usepackage{jheppub}
\usepackage{tikz}
\usepackage{subcaption}
\usepackage{simpler-wick}
\usepackage{bbm}

\DeclareMathOperator{\Tr}{Tr}

\newcommand{\ri}{\mathrm{i}}
\renewcommand{\th}{\theta}

\newcommand{\cob}{\delta}

\newcommand{\hf}{\frac{1}{2}}

\newcommand{\til}[1]{\widetilde{#1}}

\renewcommand{\b}[1]{\overline{#1}}

\newcommand{\lap}{\Delta}
\newcommand{\bra}{\langle}
\newcommand{\ket}{\rangle}
\newcommand{\la}{\lambda}

\newcommand{\h}[1]{\widehat{#1}}

\newcommand{\rt}[1]{\sqrt{#1}}
\newcommand{\cO}{\mathcal{O}}

\newcommand{\cH}{\mathcal{H}}

\newcommand{\cC}{\mathcal{C}}

\newcommand{\cE}{\mathcal{E}}

\newcommand{\id}{\mathbbm{1}}

\makeatletter
\gdef\@fpheader{}
\makeatother
\begin{document}
\title{More on doubled Hilbert space in double-scaled SYK}

\author{Kazumi Okuyama}

\affiliation{Department of Physics, 
Shinshu University, 3-1-1 Asahi, Matsumoto 390-8621, Japan}

\emailAdd{kazumi@azusa.shinshu-u.ac.jp}

\abstract{
We continue our study of the doubled Hilbert space
formalism of double-scaled SYK model
initiated in [arXiv:2401.07403].
We show that the 1-particle Hilbert space introduced by Lin and Stanford
in [arXiv:2307.15725]
is related to the doubled Hilbert space 
$\mathcal{H}_0\otimes \mathcal{H}_0$ of
the 0-particle Hilbert space $\mathcal{H}_0$
by some linear isomorphism $\mathcal{C}$. 
It turns out that the entangler and disentangler
appeared in our previous study are given by
the dual (or adjoint) of $\mathcal{C}^{-1}$ and $\mathcal{C}$.
}

\maketitle

\section{Introduction}\label{sec:intro}
In the AdS/CFT correspondence \cite{Maldacena:1997re,Witten:1998qj}, 
it is expected that the Hilbert space of
the boundary theory is isomorphic to the bulk Hilbert space of
quantum gravity.
However, in general it is difficult to check this correspondence 
since the Hilbert space of
bulk quantum gravity is not well-understood.
Recently, it is argued in \cite{Lin:2022rbf}
that the double-scaled SYK (DSSYK) model 
\cite{Cotler:2016fpe,Berkooz:2018jqr}
is a useful toy model to study the bulk Hilbert space of quantum gravity.
As shown in \cite{Berkooz:2018jqr}, the computation of the correlators in DSSYK
reduces to a counting problem of chord diagrams.
In the absence of matter particles, 
the bulk Hilbert space of the holographic dual of DSSYK
is identified with the Fock space of $q$-deformed oscillator
$A_\pm$ which creates/annihilates the chords.
In \cite{Lin:2023trc},
the bulk Hilbert space 
in the presence of matter particles
is proposed by using the structure of the coproduct
of $q$-deformed oscillators
to define the ``chord algebra'' 
acting on the multi-particle states. 

In our previous paper \cite{Okuyama:2024yya},
we developed a formalism of doubled Hilbert space
to write down the matter correlators of DSSYK in a simple form.
We found that the result of matter correlators
in \cite{Berkooz:2018jqr} can be recast into the form of the overlap of the vacuum 
state and an entangled state in the doubled Hilbert space, 
with some insertion of the operator which counts the intersection of chords.  
It turned out 
that this intersection-counting operator
should be conjugated by certain ``entangler'' and ``disentangler'' \cite{Okuyama:2024yya}.

In a recent paper \cite{Xu:2024hoc}, it was shown that
the 1-particle state
in \cite{Lin:2023trc} is written as a linear combination
of the states in the doubled Hilbert space 
considered in \cite{Okuyama:2024yya}.\footnote{The existence 
of such a relation has been suggested in \cite{Stanford-talk}.}
This is essentially equivalent to the expansion of the
bivariate $q$-Hermite polynomial
in terms of the bi-linear combination 
of the ordinary $q$-Hermite polynomials
\cite{riley2021bivariate}.
Thus, the result of \cite{Xu:2024hoc} explains the 
relation between the two different approaches in \cite{Lin:2023trc}
and \cite{Okuyama:2024yya}.
However, there remain some unanswered questions:
\renewcommand{\theenumi}{Q\arabic{enumi}}
\renewcommand{\labelenumi}{(\theenumi)}
\begin{enumerate}
\item How is the coproduct of $q$-deformed oscillators in
\cite{Lin:2023trc} realized in the doubled Hilbert space in \cite{Okuyama:2024yya}?
\label{item:Q1}
\item What is the analogue of the entangler and the disentangler
in the formalism of \cite{Lin:2023trc}?
\label{item:Q2}
\end{enumerate}
In this paper, we will give a partial answer to \eqref{item:Q1} and
a complete answer to \eqref{item:Q2}.
The result of \cite{Xu:2024hoc} implies that
the 1-particle Hilbert space $\cH_1$
in \cite{Lin:2023trc} is isomorphic to the 
doubled Hilbert space $\cH_0\otimes\cH_0$ 
considered in \cite{Okuyama:2024yya}
\begin{equation}
\begin{aligned}
 \cH_1\cong\cH_0\otimes\cH_0,
\end{aligned} 
\end{equation}
where $\cH_0$ is the 0-particle Hilbert space, i.e. the Fock space of $q$-deformed oscillator $A_\pm$.
We find that this isomorphism 
is realized by a linear operator $\cC$ in \eqref{eq:C-def}
\begin{equation}
\begin{aligned}
 \cC: \cH_0\otimes\cH_0\to\cH_1.
\end{aligned} 
\label{eq:map-C}
\end{equation}
We find that the coproduct of 
$q$-deformed oscillators appears from the direct product 
$A_\pm\otimes\id$ and $\id\otimes A_\pm$ after performing 
the conjugation by $\cC$,
up to a certain rearrangement (see \eqref{eq:rearrange}). 
The entangler and disentangler appeared in \cite{Okuyama:2024yya}
are given by the dual (or adjoint)
of $\cC^{-1}$ and $\cC$, respectively. 

This paper is organized as follows.
In section \ref{sec:review}, we briefly review the result of DSSYK in 
\cite{Berkooz:2018jqr} and introduce the 0-particle Hilbert space as the
Fock space of $q$-deformed oscillator $A_\pm$.
In section \ref{sec:1particle}, after reviewing the result of \cite{Xu:2024hoc,riley2021bivariate}, we find the isomorphism $\cC$ between the state in 
\cite{Lin:2023trc} and the direct product state in $\cH_0\otimes\cH_0$.
In subsection \ref{sec:coproduct}, we give a partial answer to \eqref{item:Q1}.
In subsection \ref{sec:inner}, we compute the inner product 
on the 1-particle Hilbert space. In subsection \ref{sec:entangler},
we give an answer to \eqref{item:Q2}.
Finally,
we conclude in section \ref{sec:conclusion} with some discussion
on the future problems.

\section{Review of the 0-particle Hilbert space}\label{sec:review}
In this section we briefly review the 
0-particle Hilbert space
of DSSYK \cite{Berkooz:2018jqr}.
SYK model is defined by the Hamiltonian for 
$N$ Majorana fermions $\psi_i~(i=1,\cdots,N)$
obeying $\{\psi_i,\psi_j\}=2\cob_{i,j}$
with all-to-all $p$-body interaction
\begin{equation}
\begin{aligned}
 H=\ri^{p/2}\sum_{1\leq i_1<\cdots<i_p\leq N}
J_{i_1\cdots i_p}\psi_{i_1}\cdots\psi_{i_p},
\end{aligned} 
\end{equation}
where $J_{i_1\cdots i_p}$ is a random coupling drawn from the Gaussian distribution.
DSSYK is defined by the scaling limit
\begin{equation}
\begin{aligned}
 N,p\to\infty\quad\text{with}\quad \la=\frac{2p^2}{N}:\text{fixed}.
\end{aligned} 
\label{eq:scaling}
\end{equation}
As shown in \cite{Berkooz:2018jqr}, the ensemble average of the moment $\Tr H^k$ 
reduces to a counting problem
of the intersection number of chord diagrams, which in turn 
is solved by introducing the transfer matrix $T$
\begin{equation}
\begin{aligned}
 \bra \Tr H^k\ket_J=\sum_{\text{chord diagrams}}q^{\#(\text{intersections})}
=\bra 0|T^k|0\ket
\end{aligned}
\label{eq:moment} 
\end{equation}
where $q=e^{-\la}$
and $T$ is given by the 
$q$-deformed oscillator $A_{\pm}$
\begin{equation}
\begin{aligned}
 T=\frac{A_{+}+A_{-}}{\rt{1-q}}.
\end{aligned} 
\end{equation}
$A_{\pm}$ obey the commutation relation\footnote{Usually, the $q$-deformed oscillator
is defined by the relation
$aa^\dag-qa^\dag a=1$.
Our $A_\pm$ is related to such $a,a^\dag$ by a rescaling
\begin{equation}
\begin{aligned}
 A_{-}=a\rt{1-q},\quad A_{+}=a^\dag\rt{1-q}.
\end{aligned} 
\end{equation}
}
\begin{equation}
\begin{aligned}
 A_{-}A_{+}-qA_{+} A_{-}=1-q,
\end{aligned} 
\end{equation}
and they act on the chord number state $|n\ket$ as
\footnote{Our $|n\ket$ differs from the state
used in \cite{Berkooz:2018jqr} by an $n$-dependent normalization constant.
See e.g. \cite{Okuyama:2022szh} for the precise relation.}
\begin{equation}
\begin{aligned}
 A_{+}|n\ket=\rt{1-q^{n+1}}|n+1\ket,\quad
A_{-}|n\ket=\rt{1-q^{n}}|n-1\ket.
\end{aligned} 
\label{eq:Apm-n}
\end{equation}
Here $|n\ket$ denotes
the state with $n$ chords 
\begin{equation}
\begin{aligned}
 \begin{tikzpicture}[scale=0.75]
\draw (0,-1)--(0,1);
\draw (0.5,-1)--(0.5,1);
\draw (1,-1)--(1,1);
\draw (1.5,-1)--(1.5,1);
\draw (2,-1)--(2,1);
\draw (2.5,-1)--(2.5,1);
\draw (3,-1)--(3,1);
\draw[dashed] (-0.5,0)--(3.5,0);
\draw (-1,0) node [left]{$|n\ket=$};
\draw (1.5,1) node [above] {$\overbrace{~\hskip23mm~}^{}$};
\draw (1.5,1.5) node [above] {$n~\text{chords}$};
\end{tikzpicture}
\end{aligned} 
\label{eq:def-n}
\end{equation}
and the dashed line in \eqref{eq:def-n} represents
a constant (Euclidean) time slice.
From \eqref{eq:Apm-n}, it follows that
$A_\pm$ satisfy
\begin{equation}
\begin{aligned}
 A_{+} A_{-}=1-q^{\h{N}},\quad
A_{-}A_{+}=1-q^{\h{N}+1},
\end{aligned} 
\label{eq:AA-N}
\end{equation}
where $\h{N}$ is the number operator
\begin{equation}
\begin{aligned}
 \h{N}|n\ket=n|n\ket.
\end{aligned} 
\end{equation}
From \eqref{eq:Apm-n}, one can show that $|n\ket$ is written as
\begin{equation}
\begin{aligned}
 |n\ket=\frac{A_+^n}{\rt{(q;q)_n}}|0\ket,
\end{aligned} 
\label{eq:n-state}
\end{equation}
where $(q;q)_n$ denotes the $q$-Pochhammer symbol
\begin{equation}
\begin{aligned}
 (a;q)_n=\prod_{k=0}^{n-1}(1-aq^k).
\end{aligned} 
\end{equation}
In what follows, we also use the notation
\begin{equation}
\begin{aligned}
 (a_1,\cdots,a_s;q)_\infty&=\prod_{j=1}^s(a_j;q)_\infty,\\
(ae^{\pm\ri\th};q)_\infty&=(ae^{\ri\th},ae^{-\ri\th};q)_\infty.
\end{aligned} 
\end{equation}

The Hamiltonian $A_{-}+A_{+}$ can be diagonalized 
by the $|\th\ket$ basis and the eigenvalue is given by
$2\cos\th$
\begin{equation}
\begin{aligned}
 (A_{-}+A_{+})|\th\ket=2\cos\th|\th\ket,\qquad
\bra\th|(A_{-}+A_{+})=2\cos\th\bra\th|.
\end{aligned} 
\label{eq:eigen-th}
\end{equation}
The overlap of $\bra\th|$ and $|n\ket$  is
given by the $q$-Hermite polynomial $H_n(x|q)$
with $x=\cos\th$
\begin{equation}
\begin{aligned}
 \bra \th|n\ket=\bra n|\th\ket=\frac{H_{n}(\cos\th|q)}{\rt{(q;q)_n}}.
\end{aligned} 
\end{equation}
From the orthogonality of the $q$-Hermite polynomials, one can show that
\begin{equation}
\begin{aligned}
  \bra n|m\ket=\int_0^\pi\frac{d\th}{2\pi}\mu(\th)\bra n|\th\ket\bra \th|m\ket
=\cob_{n,m},
\end{aligned} 
\label{eq:cob-nm}
\end{equation}
where the measure factor $\mu(\th)$ is given by
\begin{equation}
\begin{aligned}
 \mu(\th)=(q,e^{\pm2\ri\th};q)_\infty.
\end{aligned} 
\end{equation}

To summarize, the 
0-particle Hilbert space $\cH_0$
of DSSYK, i.e. the Hilbert space without a matter particle, is given by the Fock
space of $q$-deformed oscillator $A_{\pm}$ and $\cH_0$ is spanned by
the chord number states $\{|n\ket\}_{n=0,1,\cdots}$
\begin{equation}
\begin{aligned}
 \cH_0=\bigoplus_{n=0}^\infty\mathbb{C}|n\ket.
\end{aligned} 
\end{equation}
In the next section, we will consider the 1-particle Hilbert space
$\cH_1$ of DSSYK.

\section{The 1-particle Hilbert space of DSSYK}\label{sec:1particle}
As shown in \cite{Berkooz:2018jqr}, one can introduce the matter operator 
$\cO_\lap$ in DSSYK
\begin{equation}
\begin{aligned}
 \cO_\lap=\ri^{s/2}\sum_{1\leq i_1<\cdots< i_s\leq N}K_{i_1\cdots i_s}\psi_{i_1}\cdots\psi_{i_s}
\end{aligned} 
\label{eq:matter-M}
\end{equation}
with a Gaussian random coefficient $K_{i_1\cdots i_s}$
which is drawn independently from the random coupling
$J_{i_1\cdots i_p}$ in the SYK Hamiltonian. The number $s$ of fermions in
\eqref{eq:matter-M} is related to the dimension $\lap$ of the operator
$\cO_\lap$ by
\begin{equation}
\begin{aligned}
 \lap=\frac{s}{p}.
\end{aligned} 
\end{equation} 
The correlator of matter operators 
can be also written as a  
counting problem of the chord diagrams
\begin{equation}
\begin{aligned}
 \sum_{\text{chord diagrams}}q^{\#(H\text{-}H\,\text{intersections})}
q^{\lap_i\#(H\text{-}\cO_{\lap_i}\,\text{intersections})}
q^{\lap_i\lap_j\#(\cO_{\lap_i}\text{-}\cO_{\lap_j}\,\text{intersections})}.
\end{aligned} 
\label{eq:chord-count}
\end{equation}
Note that there appear two types of chords in this computation: $H$-chords and 
$\cO$-chords coming from the Wick contraction of random couplings 
$J_{i_1\cdots i_p}$ and $K_{i_1\cdots i_s}$, respectively.
The $\cO$-chord is also called matter chord. 

In \cite{Okuyama:2024yya}, it is found that
the matter correlator of DSSYK can be written in a simple form
by introducing the direct product $\cH_0\otimes\cH_0$
of the 0-particle Hilbert space $\cH_0$.
The natural basis of $\cH_0\otimes\cH_0$ is given by the direct product of the 
chord number state $|n\ket\in\cH_0$
\begin{equation}
\begin{aligned}
 |n,m\ket&:=|n\ket\otimes |m\ket.
\end{aligned}
\label{eq:nm-direct} 
\end{equation}
In general, the matter correlator takes the form of $\bra0|X|0\ket$
where $X\in\text{End}(\cH_0)$ is some linear operator on $\cH_0$.
In terms of the doubled Hilbert space $\cH_0\otimes\cH_0$,
this correlator $\bra0|X|0\ket$ can be rewritten as
\begin{equation}
\begin{aligned}
 \bra0|X|0\ket=\bra 0,0|X\ket,
\end{aligned} 
\end{equation}
where the state $|X\ket\in \cH_0\otimes\cH_0$ is given by
\begin{equation}
\begin{aligned}
 |X\ket=\sum_{n,m=0}^\infty |n,m\ket\bra n|X|m\ket.
\end{aligned} 
\end{equation}

On the other hand, 
Lin and Stanford introduced the 1-particle Hilbert space $\cH_1$
in the presence of a matter operator
$\cO_\lap$ \cite{Lin:2023trc}. $\cH_1$ is spanned by the 1-particle state
\begin{equation}
\begin{aligned}
|n_L,n_R)&:=|n_L,n_R\ket^{\text{LS}}
\end{aligned}
\label{eq:LS-state} 
\end{equation}
where $n_L,n_R$ denote the number of  $H$-chords on the left and right
of the matter chord 
\begin{equation}
\begin{aligned}
 \begin{tikzpicture}[scale=0.75]
\draw (0,-1)--(0,1);
\draw (0.5,-1)--(0.5,1);
\draw (1,-1)--(1,1);
\draw (1.5,-1)--(1.5,1);
\draw (2,-1)--(2,1);
\draw (2.5,-1)--(2.5,1);
\draw (3,-1)--(3,1);
\draw[blue,thick] (3.5,-1)--(3.5,1);
\draw (4,-1)--(4,1);
\draw (4.5,-1)--(4.5,1);
\draw (5,-1)--(5,1);
\draw (5.5,-1)--(5.5,1);
\draw (6,-1)--(6,1);
\draw (6.5,-1)--(6.5,1);
\draw (7,-1)--(7,1);
\draw[dashed] (-0.5,0)--(7.5,0);
\draw (-1.2,0) node [left]{$|n_L,n_R)=$};
\draw (1.5,1) node [above] {$\overbrace{~\hskip21mm~}^{}$};
\draw (1.5,1.5) node [above] {$n_L~\text{chords}$};
\draw (5.5,1) node [above] {$\overbrace{~\hskip21mm~}^{}$};
\draw (5.5,1.5) node [above] {$n_R~\text{chords}$};
\end{tikzpicture}
\end{aligned} 
\label{eq:nm-pic}
\end{equation}
Here the middle blue line represents the matter chord. 
In order to distinguish from the direct product state
\eqref{eq:nm-direct}, we use the round bracket for the Lin-Stanford state
in \eqref{eq:LS-state}.

As discussed in \cite{Stanford-talk}, these two basis  
\eqref{eq:nm-direct} and \eqref{eq:LS-state} are related by a 
linear transformation.
It turns out that one can explicitly write down 
the Lin-Stanford state $|n_L,n_R)$ as a linear combination of 
$|n,m\ket$ in \eqref{eq:nm-direct} \cite{Xu:2024hoc,riley2021bivariate}.
To see this, it is convenient to consider the $\th$-representation of the state. 
For the state $|n,m\ket$ in \eqref{eq:nm-direct}, its $\th$-representation
is just a product of two $q$-Hermite polynomials
\begin{equation}
\begin{aligned}
 \bra\th_1,\th_2|n,m\ket=\frac{H_n(x|q)}{\rt{(q;q)_n}}\cdot
\frac{H_m(y|q)}{\rt{(q;q)_m}},
\end{aligned} 
\end{equation}
where $x=\cos\th_1,y=\cos\th_2$ and
\begin{equation}
\begin{aligned}
|\th_1,\th_2\ket=|\th_1\ket\otimes|\th_2\ket,\quad
 \bra\th_1,\th_2|=\bra\th_1|\otimes \bra\th_2|.
\end{aligned} 
\label{eq:th12-def}
\end{equation}
As shown in \cite{Xu:2024hoc}, the 
$\th$-representation of the 
Lin-Stanford state $|n,m)$ is given by the bivariate $q$-Hermite 
polynomial $H_{n,m}(x,y|q,r)$
introduced in \cite{riley2021bivariate}
\begin{equation}
\begin{aligned}
 \bra\th_1,\th_2|n,m)=
\frac{H_{n,m}(x,y|q,r)}{\rt{(q;q)_n(q;q)_m}},
\end{aligned} 
\end{equation}
where $r$ is related to the dimension $\lap$ of the matter operator $\cO_\lap$ by
\begin{equation}
\begin{aligned}
 r=q^{\lap}.
\end{aligned} 
\end{equation}
The bivariate $q$-Hermite 
polynomial $H_{n,m}(x,y|q,r)$ has the symmetry
\begin{equation}
\begin{aligned}
 H_{n,m}(x,y|q,r)=H_{m,n}(y,x|q,r)
\end{aligned} 
\end{equation}
and it satisfies the recursion relation
\begin{equation}
\begin{aligned}
 2x H_{n,m}(x,y|q,r)=&H_{n+1,m}(x,y|q,r)\\
+&(1-q^n)H_{n-1,m}(x,y|q,r)
+rq^n(1-q^m)H_{n,m-1}(x,y|q,r)
\end{aligned} 
\label{eq:rec-bivariate}
\end{equation}
with the boundary condition $H_{n,0}(x,y|q,r)=H_n(x|q)$.
This recursion relation is interpreted as the eigenvalue equation
for the left Hamiltonian $H_L=A_L+A_L^\dag$ \cite{Lin:2023trc}.
From \eqref{eq:rec-bivariate}, one can show that
the generating function of the bivariate $q$-Hermite polynomial
is given by \cite{riley2021bivariate}
\begin{equation}
\begin{aligned}
  F(s,t)=\sum_{n,m=0}^\infty\frac{s^nt^m}{(q;q)_n(q;q)_m}H_{n,m}(x,y|q,r)
=\frac{(rst;q)_\infty}{(se^{\pm\ri\th_1},te^{\pm\ri\th_2};q)_\infty}.
\end{aligned} 
\label{eq:gen-Hnm}
\end{equation}
On the other hand, the generating function
of the $q$-Hermite polynomial is given by
\begin{equation}
\begin{aligned}
 \sum_{n=0}^\infty\frac{s^n}{(q;q)_n}H_n(\cos\th|q)
=\frac{1}{(se^{\pm\ri\th};q)_\infty}.
\end{aligned} 
\label{eq:gen-Hn}
\end{equation}
Using the identity
\begin{equation}
\begin{aligned}
 (a;q)_\infty=\sum_{k=0}^\infty \frac{(-a)^k q^{\hf k(k-1)}}{(q;q)_k}
\end{aligned} 
\end{equation}
and \eqref{eq:gen-Hn},  one can easily see from \eqref{eq:gen-Hnm} that
the bivariate $q$-Hermite polynomial is written
as a bi-linear combination of the $q$-Hermite polynomials
\begin{equation}
\begin{aligned}
 H_{n,m}(x,y|q,r)=\sum_{k=0}^{\min(n,m)}\frac{(-r)^k q^{\hf k(k-1)}(q;q)_n(q;q)_m}
{(q;q)_k (q;q)_{n-k}(q;q)_{m-k}} H_{n-k}(x|q)H_{m-k}(y|q).
\end{aligned} 
\label{eq:bi-lin}
\end{equation}
In terms of the states, \eqref{eq:bi-lin} is written as
\begin{equation}
\begin{aligned}
 |n,m)=\sum_{k=0}^{\min(n,m)}\frac{(-r)^k q^{\hf k(k-1)}}{(q;q)_k}
\rt{\frac{(q;q)_n(q;q)_m}{(q;q)_{n-k}(q;q)_{m-k}}}|n-k,m-k\ket.
\end{aligned} 
\label{eq:lin-state}
\end{equation}
Finally, using the relation
\begin{equation}
A_{-}^k|n\ket=
\left\{
\begin{aligned}
 &\rt{\frac{(q;q)_n}{(q;q)_{n-k}}}|n-k\ket, \quad &&(k\leq n),\\
&\quad 0,\quad &&(k>n),
\end{aligned}
\right. 
\end{equation}
\eqref{eq:lin-state} 
can be recast into a very simple expression
\begin{equation}
\begin{aligned}
 |n,m)=\cC|n,m\ket
\end{aligned} 
\label{eq:state-C}
\end{equation}
where $\cC$ is given by
\begin{equation}
\begin{aligned}
 \cC=\sum_{k=0}^\infty\frac{(-r)^k q^{\hf k(k-1)}}{(q;q)_k}A_{-}^k\otimes A_{-}^k
=(rA_{-}\otimes A_{-};q)_\infty.
\end{aligned} 
\label{eq:C-def}
\end{equation}
Namely, as advertised in \eqref{eq:map-C},
the doubled Hilbert space $\cH_0\otimes\cH_0$ in \cite{Okuyama:2024yya}
and the 1-particle Hilbert space $\cH_1$ in \cite{Lin:2023trc}
are related by the
linear transformation $\cC$.

\subsection{Chord algebra and conjugation by $\cC$}\label{sec:coproduct}
We have seen that the Lin-Stanford state 
$|n,m)$ is obtained from the direct product state
$|n,m\ket$ by acting the linear operator $\cC$ in \eqref{eq:C-def}.
Thus, we can naturally define the action of the $q$-deformed oscillator
on the 1-particle state $|n,m)$ by performing the conjugation with respect to
$\cC$.
It is obvious that $A_{-}\otimes\id$ and $\id\otimes A_{-}$ commute with $\cC$
\begin{equation}
\begin{aligned}
 \cC^{-1}(A_{-}\otimes\id)\cC=A_{-}\otimes\id,\quad
\cC^{-1}(\id\otimes A_{-})\cC=\id\otimes A_{-}.
\end{aligned} 
\label{eq:CA-C}
\end{equation}
On the other hand, 
$A_{+}\otimes\id$ and $\id\otimes A_{+}$ have a nontrivial commutation relation 
with $\cC$.
Let us consider the 
commutator of
$A_{+}\otimes\id$ and $\cC$
\begin{equation}
\begin{aligned}
 [A_{+}\otimes\id, \cC]&=
\sum_{k=0}^\infty \frac{(-r)^k q^{\hf k(k-1)}}{(q;q)_k}[A_+,A_{-}^k]\otimes A_{-}^k.
\end{aligned}
\label{eq:commAC}  
\end{equation}
From \eqref{eq:AA-N} we find
\begin{equation}
\begin{aligned}
 [A_+,A_{-}^k]&=(1-q^{\h{N}})A_{-}^{k-1}-A_{-}^{k-1}(1-q^{\h{N}+1})\\
&=-q^{\h{N}}A_{-}^{k-1}+A_{-}^{k-1}q^{\h{N}+1}.
\end{aligned}
\label{eq:commAA} 
\end{equation}
Using $q^{\h{N}}A_{-}^{k-1}=q^{1-k}A_{-}^{k-1}q^{\h{N}}$, 
\eqref{eq:commAA} becomes
\begin{equation}
\begin{aligned}
 [A_+,A_{-}^k]&=-(1-q^k)q^{1-k}A_{-}^{k-1}q^{\h{N}}.
\end{aligned} 
\end{equation} 
Plugging this into \eqref{eq:commAC}, 
we find
\begin{equation}
\begin{aligned}
 [A_{+}\otimes\id, \cC]&=
\sum_{k=1}^\infty \frac{(-1)^{k-1}r^k q^{\hf (k-1)(k-2)}}{(q;q)_{k-1}}A_{-}^{k-1}q^{\h{N}}\otimes A_{-}^k\\
&=\cC(rq^{\h{N}}\otimes A_{-}).
\end{aligned} 
\end{equation}
Finally we arrive at the relation
\begin{equation}
\begin{aligned}
 \cC^{-1}(A_{+}\otimes\id)\cC=A_{+}\otimes\id+rq^{\h{N}}\otimes A_{-}.
\end{aligned} 
\label{eq:CA+C}
\end{equation}
Similarly, we find
\begin{equation}
\begin{aligned}
 \cC^{-1}(\id\otimes A_{+})\cC=\id\otimes A_{+}+A_{-}\otimes rq^{\h{N}}.
\end{aligned} 
\end{equation}

Using \eqref{eq:CA-C} and \eqref{eq:CA-C}, we can deduce the action
of $A_\pm\otimes\id$ on the Lin-Stanford state $|n,m)$, as follows:
\begin{equation}
\begin{aligned}
 (A_{-}\otimes \id)|n,m)&=(A_{-}\otimes \id)\cC|n,m\ket\\
&=\cC(A_{-}\otimes \id)|n,m\ket\\
&=\cC\bigl(\rt{1-q^n}|n-1,m\ket\bigr)\\
&=\rt{1-q^n}|n-1,m),
\end{aligned} 
\label{eq:Am-LS}
\end{equation}
and
\begin{equation}
\begin{aligned}
 (A_{+}\otimes \id)|n,m)&=(A_{+}\otimes \id)\cC|n,m\ket\\
&=\cC(A_{+}\otimes \id+rq^{\h{N}}\otimes A_{-})|n,m\ket\\
&=\cC\bigl(\rt{1-q^{n+1}}|n+1,m\ket+rq^n\rt{1-q^m}|n,m-1\ket\bigr)\\
&=\rt{1-q^{n+1}}|n+1,m)+rq^n\rt{1-q^m}|n,m-1).
\end{aligned}
\label{eq:Ap-LS} 
\end{equation}
Using \eqref{eq:eigen-th}, \eqref{eq:Am-LS},
and \eqref{eq:Ap-LS},
the recursion relation \eqref{eq:rec-bivariate}
of the bivariate $q$-Hermite polynomials 
is obtained by evaluating 
$\bra\th_1,\th_2|\bigl[(A_{-}+A_{+})\otimes\id\bigr]|n,m)$.
The action of $\id\otimes A_\pm$ on the Lin-Stanford state $|n,m)$
can be found in a similar manner.

In \cite{Lin:2023trc}, the chord algebra on the 1-particle
Hilbert space is introduced by defining the coproduct of $q$-deformed oscillator
$A_\pm$
\begin{equation}
\begin{aligned}
A_L&=A_{-}\otimes\id+rq^{\h{N}}\otimes A_{-},\quad
&A_R&=\id\otimes A_{-}+A_{-}\otimes rq^{\h{N}},\\
 A_L^\dag&=A_{+}\otimes \id,\quad
&A_R^\dag&=\id\otimes A_{+}.
\end{aligned} 
\label{eq:ALAR-def}
\end{equation}
They satisfy the relations
\begin{equation}
\begin{aligned}
 &A_LA_L^\dag-qA_L^\dag A_L=A_RA_R^\dag-qA_R^\dag A_R=1-q,\\
&[A_L,A_R]=[A_L^\dag,A_R^\dag]=0,\\
&[A_L,A_R^\dag]=[A_R,A_L^\dag]=(1-q)rq^{\h{N}}\otimes q^{\h{N}}.
\end{aligned} 
\label{eq:copro}
\end{equation}
We should stress that $A_L,A_L^\dag$ are not equal to 
$\cC^{-1}(A_{\pm}\otimes \id)\cC$. Similarly, 
$A_R,A_R^\dag$ are not equal to 
$\cC^{-1}(\id\otimes A_{\pm})\cC$
\begin{equation}
\begin{aligned}
 A_L &\ne \cC^{-1}(A_{-}\otimes \id)\cC,\quad
&A_R&\ne \cC^{-1}(\id\otimes A_{-})\cC,\\
A_L^\dag &\ne \cC^{-1}(A_{+}\otimes \id)\cC,\quad
&A_R^\dag &\ne \cC^{-1}(\id\otimes A_{+})\cC.
\end{aligned} 
\end{equation}
In fact, had we defined $A_L$ and $A_R^\dag$ by
$ A_L=\cC^{-1}(A_{-}\otimes \id)\cC,
A_R^\dag=\cC^{-1}(\id\otimes A_{+})\cC$,
they would commute with each other $[A_L,A_R^\dag]=0$, which differs
from the nontrivial coproduct structure in the last line of \eqref{eq:copro}.
The  coproduct in \cite{Lin:2023trc} arises
after rearranging (or twisting) $\cC^{-1}(A_{\pm}\otimes \id)\cC$
into the creation and annihilation parts
\begin{equation}
\left\{
\begin{aligned}
 &\cC^{-1}(A_{-}\otimes\id)\cC=\textcolor{blue}{A_{-}\otimes\id} \\
&\cC^{-1}(A_{+}\otimes\id)\cC=\textcolor{red}{A_{+}\otimes\id} +\textcolor{blue}{rq^{\h{N}}\otimes A_{-}}
\end{aligned} 
\right.\quad
\Rightarrow\quad
\left\{
\begin{aligned}
 &A_L=\textcolor{blue}{A_{-}\otimes\id+ rq^{\h{N}}\otimes A_{-}}\\
&A_L^\dag=\textcolor{red}{A_{+}\otimes\id}
\end{aligned} 
\right.
\label{eq:rearrange}
\end{equation}
Note that the Hamiltonian $A_{-}+A_{+}$ is mapped to $H_L=A_L+A_L^\dag$
and $H_R=A_R+A_R^\dag$
by a simple conjugation by $\cC$
\begin{equation}
\begin{aligned}
 \cC^{-1}\bigl[(A_{-}+A_{+})\otimes \id\bigr]\cC&=A_L+A_L^\dag,\\
\cC^{-1}\bigl[\id\otimes (A_{-}+A_{+})\bigr]\cC&=A_R+A_R^\dag.
\end{aligned} 
\label{eq:map-HLR}
\end{equation}
The rearrangement (or twist) in \eqref{eq:rearrange}
is different from the so-called Drinfeld twist.
It would be interesting to understand the physical origin of the coproduct in 
\cite{Lin:2023trc}.

\subsection{Inner product of 1-particle states}\label{sec:inner}
In this subsection, we consider the inner product on the 
1-particle Hilbert space $\cH_1$.
For the doubled Hilbert space $\cH_0\otimes\cH_0$,
it is natural to define
the direct product of inner product in
\eqref{eq:cob-nm}
\begin{equation}
\begin{aligned}
 \bra \psi_1|\psi_2\ket=\int_0^\pi\prod_{i=1,2}\frac{d\th_i}{2\pi}\mu(\th_i)
\bra \psi_1|\th_1,\th_2\ket\bra\th_1,\th_2|\psi_2\ket.
\end{aligned} 
\label{eq:inner-direct}
\end{equation}
For the state $|n,m\ket$, we find
\begin{equation}
\begin{aligned}
 \bra n,m|n',m'\ket=\cob_{n,n'}\cob_{m,m'}.
\end{aligned} 
\label{eq:inner-direct-nm}
\end{equation}
One can easily show that the adjoint of $A_\pm\otimes \id$
is $A_\mp\otimes \id$ with respect to this inner product \eqref{eq:inner-direct}
\begin{equation}
\begin{aligned}
 (A_\pm\otimes \id)^\dag=A_\mp\otimes \id.
\end{aligned} 
\end{equation}
Similarly, we find $(\id\otimes A_\pm)^\dag=\id\otimes A_\mp$.

On the other hand, the inner product on the 1-particle 
Hilbert space
is defined by \cite{Xu:2024hoc}
\begin{equation}
\begin{aligned}
 (\psi_1|\psi_2)=
\int_0^\pi\prod_{i=1,2}\frac{d\th_i}{2\pi}\mu(\th_i)\bra\th_1|r^{\h{N}}|\th_2\ket
(\psi_1|\th_1,\th_2\ket\bra\th_1,\th_2|\psi_2),
\end{aligned} 
\label{eq:inner-LS}
\end{equation}
where the factor $\bra\th_1|r^{\h{N}}|\th_2\ket$
is the contribution of a matter particle
\begin{equation}
\begin{aligned}
 \bra\th_1|r^{\h{N}}|\th_2\ket=\sum_{n=0}^\infty
\frac{r^n}{(q;q)_n}H_n(\cos\th_1|q)
H_n(\cos\th_2|q)=\frac{(r^2;q)_\infty}{(re^{\ri(\pm\th_1\pm\th_2)};q)_\infty}.
\end{aligned} 
\end{equation}
In \cite{Lin:2023trc}, it is argued 
that $A_L,A_L^\dag$ in \eqref{eq:ALAR-def} are adjoint with each other with
respect to the inner product in \eqref{eq:inner-LS}.
It turns out that, to define the well-behaved operators we have to conjugate 
$A_L,A_L^\dag$ in \eqref{eq:ALAR-def} by $\cC$
\begin{equation}
\begin{aligned}
 \til{A}_L&=\cC A_L\cC^{-1}=\cC(A_{-}\otimes \id+rq^{\h{N}}\otimes A_{-})\cC^{-1},\\
\til{A}_L^\dag&=
\cC A_L^\dag \cC^{-1}=\cC(A_{+}\otimes\id)\cC^{-1}.
\end{aligned} 
\label{tilAL}
\end{equation}
These operators act on the Lin-Stanford state $|n,m)$ as
\begin{equation}
\begin{aligned}
 \til{A}_L|n,m)&=\rt{1-q^n}|n-1,m)+rq^n\rt{1-q^m}|n,m-1),\\
\til{A}_L^\dag|n,m)&=\rt{1-q^{n+1}}|n+1,m).
\end{aligned}
\label{eq:ALact} 
\end{equation}

We can prove that the adjoint of $\til{A}_L$ is $\til{A}_L^\dag$  
with
respect to the inner product in \eqref{eq:inner-LS}, as follows.
To this end, it is convenient to work with the generating function
\begin{equation}
\begin{aligned}
 |\psi(s,t))=\sum_{n,m=0}^\infty\frac{s^nt^m}{\rt{(q;q)_n(q;q)_m}}|n,m),
\end{aligned} 
\end{equation}
whose $\th$-representation is the generating function $F(s,t)$
in \eqref{eq:gen-Hnm}
\begin{equation}
\begin{aligned}
 \bra\th_1,\th_2|\psi(s,t))=F(s,t).
\end{aligned} 
\end{equation}
From \eqref{eq:ALact}, 
one can easily show that
\begin{equation}
\begin{aligned}
\til{A}_L|\psi(s,t))&=s|\psi(s,t))+rt|\psi(qs,t)),\\
 \til{A}_L^\dag |\psi(s,t))&=\frac{1}{s}|\psi(s,t))-
\frac{1}{s}|\psi(qs,t)).
\end{aligned} 
\label{eq:AL-psi}
\end{equation}
The inner product of $|\psi(s,t))$ with respect to 
\eqref{eq:inner-LS} is given by
\begin{equation}
\begin{aligned}
  (\psi(s,t)|\psi(s',t'))&=
\int_0^\pi\prod_{i=1,2}\frac{d\th_i}{2\pi}\mu(\th_i)
\frac{(r^2;q)_\infty}{(re^{\ri(\pm\th_1\pm\th_2)};q)_\infty}
\frac{(rst,rs't';q)_\infty}{(se^{\pm\ri \th_1},s'e^{\pm\ri \th_1},te^{\pm\ri \th_2},t'e^{\pm\ri \th_2};q)_\infty}.
\end{aligned} 
\label{eq:inner-psi-int}
\end{equation}
Using the Askey-Wilson integral \cite{Askey1985SomeBH}
\begin{equation}
\begin{aligned}
I(a_1,a_2,a_3,a_4)= \int_0^\pi\frac{d\th}{2\pi}\mu(\th)\prod_{i=1}^4\frac{1}{(a_ie^{\pm\ri\th};q)_\infty}=\frac{\prod_i(a_i;q)_\infty}{\prod_{i<j}(a_ia_j;q)_\infty},
\end{aligned} 
\end{equation}
\eqref{eq:inner-psi-int}
can be evaluated by integrating over $\th_2$ first,
 and then integrating over $\th_1$
\begin{equation}
\begin{aligned}
 (\psi(s,t)|\psi(s',t'))&=
\int_0^\pi\frac{d\th_1}{2\pi}\mu(\th_1)\frac{(r^2,rst,rs't';q)_\infty}{(se^{\pm\ri \th_1},s'e^{\pm\ri \th_1};q)_\infty}
I(re^{\ri\th_1},re^{-\ri\th_1},t,t')\\
&=\int_0^\pi\frac{d\th_1}{2\pi}\mu(\th_1)\frac{(rst,rs't',r^2tt';q)_\infty}{(se^{\pm\ri \th_1},s'e^{\pm\ri \th_1},rte^{\pm\ri\th_1},rt'e^{\pm\ri\th_1},tt';q)_\infty}\\
&=\frac{(r^2ss'tt';q)_\infty}{(ss',tt',rst',rs't;q)_\infty}.
\end{aligned} 
\label{eq:inner-psi}
\end{equation}
Finally, using \eqref{eq:AL-psi} and 
\eqref{eq:inner-psi} we find
\begin{equation}
\begin{aligned}
 (\til{A}_L\psi(s,t)|\psi(s',t'))=(\psi(s,t)|\til{A}_L^\dag \psi(s',t')),
\end{aligned} 
\label{eq:adj-AL}
\end{equation}
which means that the adjoint of 
$\til{A}_L$ is equal to $\til{A}_L^\dag$ 
with respect to the inner product \eqref{eq:inner-LS}.
This property was explained in \cite{Lin:2023trc} by a combinatorial argument.
Here we have seen that \eqref{eq:adj-AL} can be proved algebraically by using
the inner product of the generating functions.

\subsection{$R$-matrix and entangler/disentangler}\label{sec:entangler}
Let us consider the inner product of the $\th$-basis
in the presence of the matter contribution $\bra\th_1|r^{\h{N}}|\th_2\ket$.
We define
\begin{equation}
\begin{aligned}
 |\th_1,\th_2)=|\th_1\ket\otimes|\th_2\ket.
\end{aligned} 
\end{equation}
Here the round bracket of $|\th_1,\th_2)$ is meant to indicate that
the inner product is computed by the definition in \eqref{eq:inner-LS}.
Using the relation
\begin{equation}
\begin{aligned}
 \bra\th|\th'\ket=\frac{2\pi\cob(\th-\th')}{\mu(\th)},
\end{aligned} 
\end{equation}
we find
\begin{equation}
\begin{aligned}
 (\th_1,\th_2|\th_1',\th_2')=\bra\th_1|r^{\h{N}}|\th_2\ket
\bra\th_1|\th_1'\ket\bra\th_2|\th_2'\ket.
\end{aligned} 
\end{equation}
This reproduces eq.(102) in \cite{Lin:2023trc}.

Next, let us consider eq.(103) in \cite{Lin:2023trc}
corresponding to the crossed four-point function of matter operators
$\cO_{\lap_1},\cO_{\lap_2}$
\begin{equation}
\begin{aligned}
 (\th_3,\th_2|q^{\lap_2\b{n}}|\th_4,\th_1)
\end{aligned} 
\label{eq:amp-4}
\end{equation}
where the inner product is defined by \eqref{eq:inner-LS} with $r=q^{\lap_1}$,
and $q^{\lap_2\b{n}}$ counts the total number of $H$-chords in the 
Lin-Stanford state
\begin{equation}
\begin{aligned}
 q^{\lap_2\b{n}}|n,m)=q^{\lap_2(n+m)}|n,m).
\end{aligned} 
\end{equation}
From \eqref{eq:state-C},
we can show that $q^{\lap_2\b{n}}$ in our notation is given by
\begin{equation}
\begin{aligned}
q^{\lap_2\b{n}}= \cC (q^{\lap_2\h{N}}\otimes q^{\lap_2\h{N}})\cC^{-1}.
\end{aligned} 
\end{equation}
From the definition of inner product
\eqref{eq:inner-LS}, \eqref{eq:amp-4} is  written
more explicitly as
\begin{equation}
\begin{aligned}
 (\th_3,\th_2|q^{\lap_2\b{n}}|\th_4,\th_1)
=\bra\th_3|q^{\lap_1\h{N}}|\th_2\ket
\bra \th_3,\th_2|
\cC (q^{\lap_2\h{N}}\otimes q^{\lap_2\h{N}})\cC^{-1}|\th_4,\th_1\ket.
\end{aligned} 
\end{equation}
Since this amplitude is real, it is equal to its complex conjugate
\begin{equation}
\begin{aligned}
(\th_3,\th_2|q^{\lap_2\b{n}}|\th_4,\th_1)&=
\bra\th_3|q^{\lap_1\h{N}}|\th_2\ket
\bigl(\bra \th_3,\th_2|
\cC (q^{\lap_2\h{N}}\otimes q^{\lap_2\h{N}})\cC^{-1}|\th_4,\th_1\ket\bigr)^\dag\\
&=\bra\th_3|q^{\lap_1\h{N}}|\th_2\ket
\bra\th_4,\th_1|(\cC^\dag)^{-1}(q^{\lap_2\h{N}}\otimes q^{\lap_2\h{N}})
\cC^\dag|\th_3,\th_2\ket.
\end{aligned} 
\label{eq:amp-Cdag}
\end{equation} 
Here the dagger of $\cC^\dag$ is defined with respect to the 
inner product \eqref{eq:inner-direct}.
This reproduces the result of $R$-matrix 
in \cite{Okuyama:2024yya} 
\begin{equation}
\begin{aligned}
 R(\th_1,\th_2,\th_3,\th_4)=
\bra\th_4,\th_1|\cE_{\lap_1}(q^{\lap_2\h{N}}\otimes q^{\lap_2\h{N}})
(\cE_{\lap_1})^{-1}|\th_3,\th_2\ket\bra\th_3|q^{\lap_1\h{N}}|\th_2\ket
\end{aligned} 
\end{equation}
under the identification
\begin{equation}
\begin{aligned}
 (\cC^\dag)^{-1}&=\frac{1}{(q^{\lap_1}A_+\otimes A_{+};q)_\infty}=\cE_{\lap_1},\\
\cC^\dag&=(q^{\lap_1}A_+\otimes A_{+};q)_\infty=(\cE_{\lap_1})^{-1}.
\end{aligned} 
\end{equation}
Namely, the entangler $\cE_{\lap_1}$ and
the disentangler $(\cE_{\lap_1})^{-1}$ in \cite{Okuyama:2024yya}
naturally arise as the adjoint of $\cC^{-1}$ and $\cC$.

\section{Conclusions and outlook}\label{sec:conclusion}
In this paper, we have studied the relation between
the 1-particle Hilbert space $\cH_1$ in \cite{Lin:2023trc}
and the doubled Hilbert space in \cite{Okuyama:2024yya}.
We found that their basis are related by the linear transformation $\cC$
in \eqref{eq:C-def}.
The entangler and the disentangler 
in the doubled Hilbert space formalism 
naturally arise as the dual (or adjoint) of $\cC^{-1}$ and $\cC$, respectively.
The left Hamiltonian $(A_{+}+A_{-})\otimes \id$
and the right Hamiltonian $\id\otimes (A_{+}+A_{-})$
in the doubled Hilbert space formalism
are mapped to $H_L$ and $H_R$ in \cite{Lin:2023trc}
by the conjugation
with respect to $\cC$ \eqref{eq:map-HLR}.
However, the left  and  right 
$q$-oscillators $A_L,A_L^\dag$ and $A_R,A_R^\dag$ 
in \cite{Lin:2023trc} are constructed by 
a nontrivial coproduct, which is obtained after a certain 
rearrangement (or twist) of the naive $\cC$-conjugation \eqref{eq:rearrange}.
It would be interesting to understand the physical meaning of 
this coproduct better.

There are many interesting open questions.
In \cite{Lin:2023trc}, the chord algebra on the multi-particle
Hilbert space was constructed 
by using the structure of coproduct successively.
It would be interesting to find a linear map
between the $k$-particle Hilbert space $\cH_k$
and the $k^{\text{th}}$ power of the 0-particle Hilbert spaces $\cH_0^{\otimes k}$
by generalizing our construction of $\cC$.
However, the multivariate $q$-Hermite polynomials have not been constructed in
the literature before, as far as we know.
This might be related to the 
difficulty of a simple characterization of the twisting 
in \eqref{eq:rearrange}. We leave this as an interesting future problem.

In general, we expect that
the bulk Hilbert space of quantum gravity has many realizations due 
to the diffeomorphism invariance of the bulk theory. 
The different realization of bulk Hilbert space corresponds to
a different way of cutting open the bulk path integral.
However, the diffeomorphism invariance implies that these 
different realizations of bulk Hilbert spaces are all isomorphic
with each other.
The isomorphism between the 1-particle Hilbert space $\cH_1$ and
the doubled Hilbert space $\cH_0\otimes\cH_0$
can be thought of as an example of such a phenomenon.
It would be interesting to learn more general lessons of 
the Hilbert space of quantum gravity
from the detailed study of DSSYK.

As discussed in \cite{Lin:2023trc},
the chord diagrams of DSSYK have a huge redundancy
and the different chord diagrams
may correspond to the same amplitude in the boundary theory.
For instance, the crossed six-point function of matter operators
has the following two different realization of the chord diagrams
(here we only draw the matter chords)
\begin{equation}
\begin{aligned}
 \begin{tikzpicture}[scale=1]
\draw[thick] (0,0) circle [radius=2];
\draw[blue,thick] (-1.41,1.41)--(1.41,-1.41); 
\draw[red,thick] (-1.41,-1.41)--(1.41,1.41); 
\draw [teal,thick] plot [smooth] coordinates {(0,2) (-1,0) (0,-2)};
\fill (1.41,1.41) circle (0.08);
\fill (-1.41,1.41) circle (0.08);
\fill (1.41,-1.41) circle (0.08);
\fill (-1.41,-1.41) circle (0.08);
\fill (0,2) circle (0.08);
\fill (0,-2) circle (0.08);
\draw (3.1,0) node [right]{$=$};
\draw[thick] (7,0) circle [radius=2];
\draw[blue,thick] (5.59,1.41)--(8.41,-1.41); 
\draw[red,thick] (5.59,-1.41)--(8.41,1.41);
\draw [teal,thick] plot [smooth] coordinates {(7,2) (8,0) (7,-2)};
\fill (5.59,1.41) circle (0.08);
\fill (5.59,-1.41) circle (0.08);
\fill (8.41,1.41) circle (0.08);
\fill (8.41,-1.41) circle (0.08);
\fill (7,2) circle (0.08);
\fill (7,-2) circle (0.08);
\end{tikzpicture}
\end{aligned}
\label{eq:G6-YB}
\end{equation}
According to the Feynman rule in \cite{Jafferis:2022wez}, we assign 
the variables
$\th_i~(i=1,\cdots,7)$ to the faces of the diagram above and
the $R$-matrix to the internal vertices. We also assign certain factors to the
vertices and edges on the boundary of the disk.
Finally we integrate over $\th_i$ with the measure 
$\int_0^\pi\prod_i \frac{d\th_i}{2\pi}\mu(\th_i)$.
It turns out that 
the chord diagrams on the left hand side and the right hand side of 
\eqref{eq:G6-YB} are actually equal thanks to 
the Yang-Baxter relation 
satisfied by the 6j-symbol of $U_q(\mathfrak{sl}_2)$ \cite{Jafferis:2022wez}.
Usually, the Yang-Baxter relation is interpreted as the integrability of the theory.
Here, the interpretation is different: the equality \eqref{eq:G6-YB}
under the Yang-Baxter move is an analogue of the bulk diffeomorphism invariance.
On the boundary theory side, the only data we can specify is the configuration of
matter operators on the boundary of the disk, and
how to draw the chord diagram on the bulk is an arbitrary choice.
This ambiguity, or redundancy, is resolved by the equality of 
the amplitude under the Yang-Baxter move, as shown in \eqref{eq:G6-YB}.

The chord diagrams in DSSYK have some reminiscence of the tensor network models
of holography \cite{Swingle:2009bg,Pastawski:2015qua,Hayden:2016cfa}.
However, there is a crucial difference:
In the chord diagram of DSSYK, the redundancy of the bulk theory is built-in
as the invariance under the Yang-Baxter move of chords and
this redundancy serves as an analogue of the bulk diffeomorphism invariance.
It would be interesting to construct a model of holographic tensor network
with this kind of redundancy.

\acknowledgments
The author would like to thank Ryo Suzuki and Takeshi Tachibana for discussion,
and Jiuci Xu for correspondence. 
This work was supported
in part by JSPS Grant-in-Aid for Transformative Research Areas (A) 
``Extreme Universe'' 21H05187 and JSPS KAKENHI 22K03594.

\bibliography{paper}
\bibliographystyle{utphys}

\end{document}